\documentstyle[epsf]{mn}
\newif\ifAMStwofonts
\AMStwofontstrue

\ifoldfss
  \ifCUPmtlplainloaded \else
    \NewTextAlphabet{textbfit} {cmbxti10} {}
    \NewTextAlphabet{textbfss} {cmssbx10} {}
    \NewMathAlphabet{mathbfit} {cmbxti10} {} 
    \NewMathAlphabet{mathbfss} {cmssbx10} {} 
  \fi
  \ifAMStwofonts
    \ifCUPmtlplainloaded \else
      \NewSymbolFont{upmath} {eurm10}
      \NewSymbolFont{AMSa} {msam10}
      \NewMathSymbol{\upi}     {0}{upmath}{19}
      \NewMathSymbol{\umu}     {0}{upmath}{16}
      \NewMathSymbol{\upartial}{0}{upmath}{40}
      \NewMathSymbol{\leqslant}{3}{AMSa}{36}
      \NewMathSymbol{\geqslant}{3}{AMSa}{3E}

    \fi
  \fi
\fi 

\ifnfssone
  \newmathalphabet{\mathit}
  \addtoversion{normal}{\mathit}{cmr}{m}{it}
  \addtoversion{bold}{\mathit}{cmr}{bx}{it}
  \newmathalphabet{\mathbfit} 
  \addtoversion{normal}{\mathbfit}{cmr}{bx}{it}
  \addtoversion{bold}{\mathbfit}{cmr}{bx}{it}
  \newmathalphabet{\mathbfss} 
  \addtoversion{normal}{\mathbfss}{cmss}{bx}{n}
  \addtoversion{bold}{\mathbfss}{cmss}{bx}{n}
  \ifAMStwofonts
    \ifCUPmtlplainloaded \else
      \UseAMStwoboldmath
      \makeatletter
      \new@mathgroup\upmath@group
      \define@mathgroup\mv@normal\upmath@group{eur}{m}{n}
      \define@mathgroup\mv@bold\upmath@group{eur}{b}{n}
      \edef\UPM{\hexnumber\upmath@group}
      \new@mathgroup\amsa@group
      \define@mathgroup\mv@normal\amsa@group{msa}{m}{n}
      \define@mathgroup\mv@bold\amsa@group{msa}{m}{n}
      \edef\AMSa{\hexnumber\amsa@group}
      \makeatother
      \mathchardef\upi="0\UPM19
      \mathchardef\umu="0\UPM16
      \mathchardef\upartial="0\UPM40
      \mathchardef\leqslant="3\AMSa36
      \mathchardef\geqslant="3\AMSa3E
    \fi
  \fi
\fi 

\ifnfsstwo
  \DeclareMathAlphabet{\mathbfit}{OT1}{cmr}{bx}{it}
  \SetMathAlphabet\mathbfit{bold}{OT1}{cmr}{bx}{it}
  \DeclareMathAlphabet{\mathbfss}{OT1}{cmss}{bx}{n}
  \SetMathAlphabet\mathbfss{bold}{OT1}{cmss}{bx}{n}
  \ifAMStwofonts
    \ifCUPmtlplainloaded \else
      \DeclareSymbolFont{UPM}{U}{eur}{m}{n}
      \SetSymbolFont{UPM}{bold}{U}{eur}{b}{n}
      \DeclareSymbolFont{AMSa}{U}{msa}{m}{n}
      \DeclareMathSymbol{\upi}{0}{UPM}{"19}
      \DeclareMathSymbol{\umu}{0}{UPM}{"16}
      \DeclareMathSymbol{\upartial}{0}{UPM}{"40}
      \DeclareMathSymbol{\leqslant}{3}{AMSa}{"36}
      \DeclareMathSymbol{\geqslant}{3}{AMSa}{"3E}
    \fi
  \fi
\fi 

\ifCUPmtlplainloaded \else
  \ifAMStwofonts \else 
    \def\upi{\pi}
    \def\umu{\mu}
    \def\upartial{\partial}
  \fi
\fi

\title{{\it COBE}-DMR constraints on the nonlinear coupling parameter:
a wavelet based method}
\author[] {L. Cay\'on$^{1}$,
E. Mart\'\i nez-Gonz\'alez$^{1}$,
F. Arg\"ueso$^{2}$, A. J. Banday$^{3}$ and K. M. G\'{o}rski$^{4,5}$.\\
1. Instituto de F\'\i sica de Cantabria, Fac. Ciencias, Av. los
	Castros s/n, 39005 Santander, Spain\\
2. Dpto. de Matem\'aticas, Universidad de Oviedo, c/ Calvo Sotelo s/n, 33007 Oviedo, Spain\\
3. Max-Planck Institut fuer Astrophysik (MPA), Karl-Schwarzschild Str.1, D-85740, Garching, Germany\\
4. European Southern Observatory (ESO), Karl-Schwarzschild Str.2,
D-85740, Garching, Germany\\
5. Warsaw University Observatory, Poland.\\}

\date{\today}

\pagerange{\pageref{firstpage}--\pageref{lastpage}}
\pubyear{2001}

\begin{document}

\maketitle

\label{firstpage}

\begin{abstract}

Nonlinearity introduced in slow-roll inflation will produce
weakly non-Gaussian CMB temperature fluctuations. We have simulated 
non-Gaussian large scale CMB maps (including {\it COBE}-DMR constraints) 
introducing an additional quadratic term in the
gravitational potential.The amount of nonlinearity being controlled by
the so called nonlinear coupling parameter $f_{nl}$. An analysis based on 
the Spherical Mexican Hat wavelet was applied to these and to 
the {\it COBE}-DMR maps.
Skewness values 
obtained at several scales were combined into a Fisher discriminant. 
Comparison of the Fisher discriminant distributions obtained for 
different nonlinear coupling parameters with the {\it COBE}-DMR values,
sets a constraint of $\vert f_{nl}\vert <1100$ at the $68\%$ confidence
level. This new constraint being tighter than the one previously
obtained by using the bispectrum by Komatsu et al. (2002).

\end{abstract}


\section{Introduction}

The {\it COBE}-DMR data have been found to be compatible with Gaussianity 
by several methods based on real (Torres et al. 1995; Kogut et al 1996; Schmalzing \& G\'{o}rski 1997; Novikov et al. 2000), spherical harmonic (Banday et al. 2000; Sandvik \& Magueijo 2000; Komatsu et al. 2002)
and wavelet space (Mukherjee, Hobson \& Lasenby 2000; 
Aghanim, Forni \& Bouchet 2000; Barreiro et al. 2000; Cay\'on et al. 2001). 
CMB observations by Boomerang, DASI and
MAXIMA-I (Netterfield et al. 2001; Pryke et al.
2001; Stompor et al. 2001) seem to indicate that any possible 
non-Gaussianity present in the 
data will more likely be produced by non-standard Inflationary models.  

It is important at the moment to study the characteristics of the 
non-Gaussianity introduced by the alternative theories to standard Inflation.
One of the simplest alternative scenarios that will generate
weakly non-Gaussian CMB fluctuations is one based on slow-roll
inflation. Investigated by
Salopek \& Bond (1990) and Gangui et al. (1994), this model generates 
non-Gaussianity in matter and radiation fluctuation fields through
features appearing in the inflaton potential. In the CMB context, the
model has been studied by Komatsu \& Spergel (2001), Verde et al. (2001) and
Komatsu et al. (2002). The gravitational potential includes now 
a quadratic term whith an amplitude regulated by the so called
non-linear coupling parameter (see Section 2).
Verde et al. (2001) conclude in their paper that CMB
data will be more sensitive to the non-Gaussianity generated by these
models than observations of high redshift objects. 
A semianalytical expression of the bispectrum generated by this model 
is presented in Komatsu \& Spergel (2001). The minimum 
non-linear coupling parameter values that will produce detectable 
non-Gaussianity (based on the bispectrum and skewness statitics) 
in different CMB data sets are estimated. These values are however 
larger than the ones expected from slow-roll inflation theories. 
Komatsu et al. (2002) revised the constraint impossed by {\it COBE}-DMR 
data on the non-linear coupling parameter based on the normalized
bispectrum. This constraint (slightly larger than the one 
estimated by Komatsu \& Spergel 2001), though very weak to say 
much about slow-roll inflation is in any case interesting in itself. 
Deviations from slow-roll inflation might include larger 
values of the non-linear coupling parameter generating non-Gaussianities
that could be detected by CMB experiments.

We want to asses in this paper which is the minimum value of the 
non-linear coupling parameter that would be constraint by the {\it COBE}-DMR 
data based on a wavelet space statistic. Comparison of this constraint
with the one obtained using the normalized bispectrum is an 
important comparison between two methods that will potentially be
used in the 
analysis of future CMB data. At the moment, simulations of CMB temperature
fluctuations generated by the slow-roll inflation model here considered are 
possible at large angular scales. 
The outline of the paper is the following. 
The simulations used in this work
are described in Section 2. The method used to constrain the
non-linear coupling parameter is based on the Fisher discriminant
built on skewness values corresponding to wavelet coefficient maps at
several scales. These method is presented in Section 3. Results are summarized 
in Section 4. Section 5 is dedicated to discussion and conclusions.

\setcounter{figure}{0}
\begin{figure*}
 \epsffile{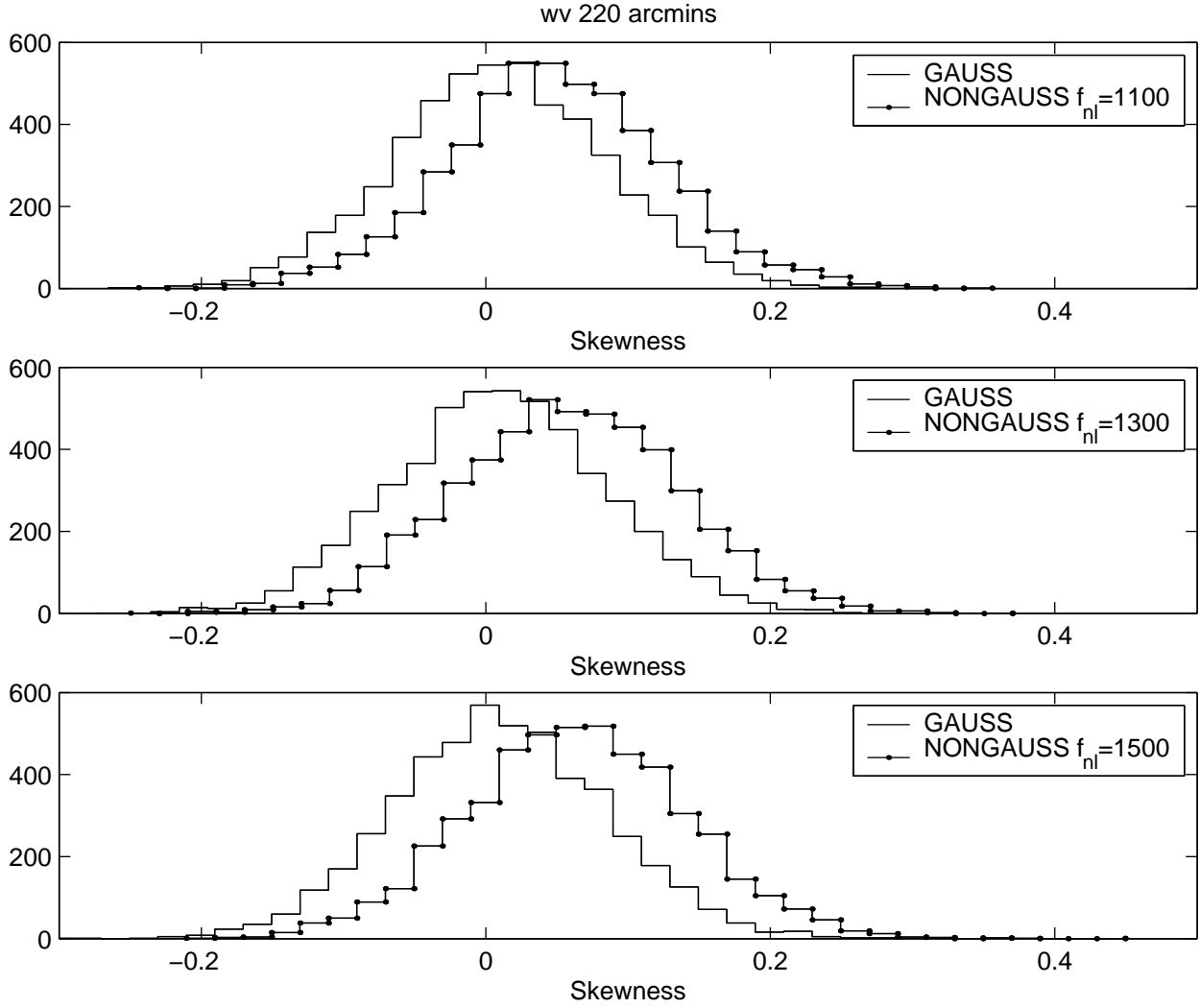}
 \caption{Distribution of skewness values obtained from simulated Gaussian
(solid line) and NonGaussian (solid-dotted line) maps after convolution with
a Spherical Mexican Hat wavelet of size 220 arcmins. Three different NonGaussian
cases are presented with $f_{nl}=1100,1300,1500$ from top to bottom
respectively}
 \label{f1}
\end{figure*}

\begin{figure*}
 \epsffile{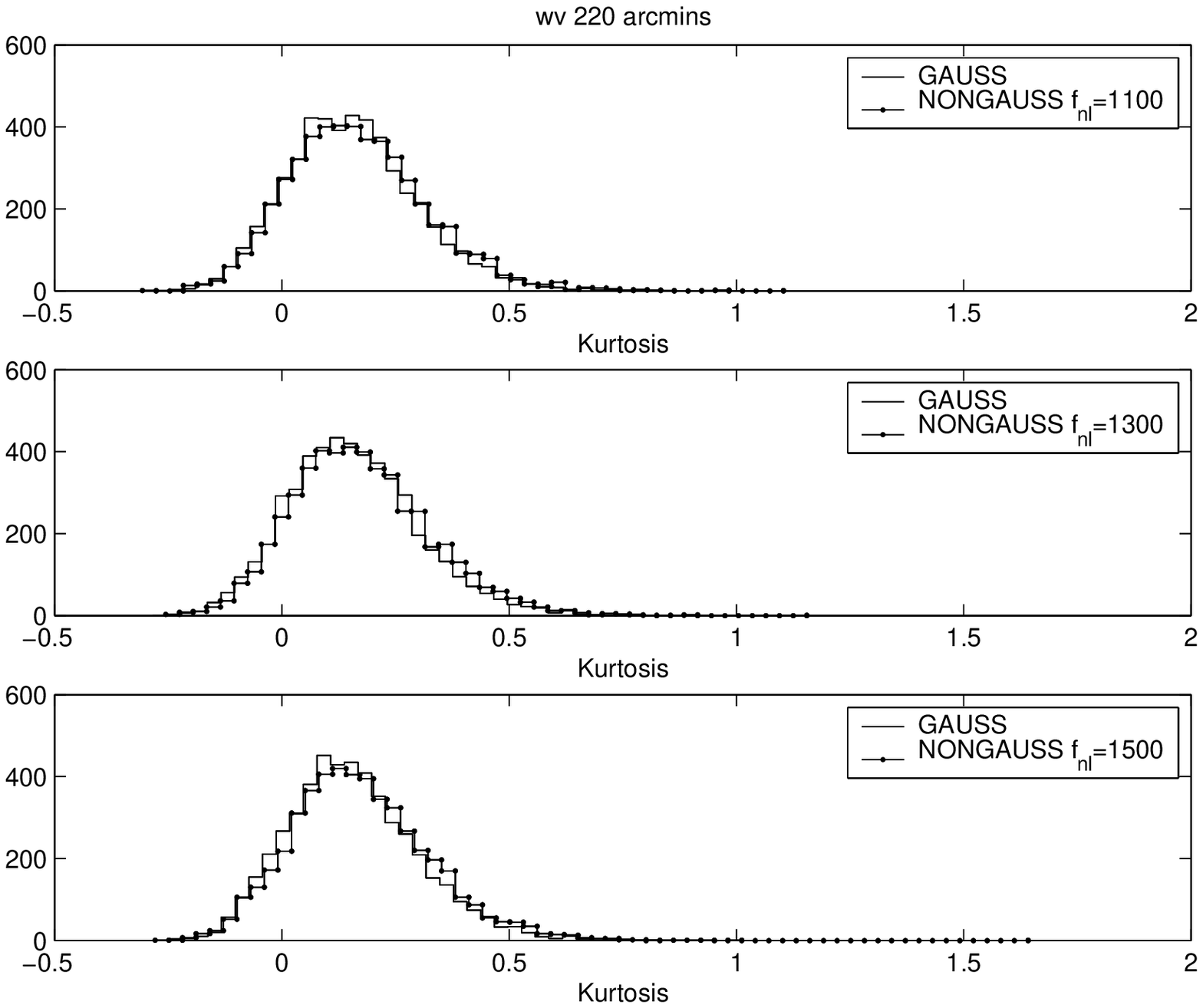}
 \caption{Distribution of kurtosis values obtained from simulated Gaussian
(solid line) and NonGaussian (solid-dotted line) maps after convolution with
a Spherical Mexican Hat wavelet of size 220 arcmins. Three different NonGaussian
cases are presented with $f_{nl}=1100,1300,1500$ from top to bottom
respectively}
 \label{f1}
\end{figure*}

\section{Simulations}

At large scales, the CMB temperature fluctuations are dominated 
by the Sachs-Wolfe effect and are therefore proportional to
the fluctuations of the gravitational potential. Weakly non-Gaussian
CMB temperature fluctuations will
be generated by the introduction of a nonlinear term in the 
gravitational potential $\phi(\vec n)$
$$\phi(\vec n)=\phi_L(\vec n)+f_{nl}(\phi_L^2(\vec n)-<\phi^2(\vec n)>),$$ 
where $\phi_L$ refers to the linear part of the gravitational potential,
$f_{nl}$ is the nonlinear coupling parameter controlling the amount of 
non-Gaussianity introduced and the brackets indicate a volume average. 
The linear part of the gravitational potential gives rise to CMB 
temperature fluctuations characterised by a power spectrum that in 
our case is assumed to be a CDM power spectrum (computed with the CMBFAST 
software) with $\Omega_b=0.05,\Omega_c=0.3,\Omega_v=0.65,\Omega_{\nu}=0,H_o=65$ km/s/Mpc and $n=1$ and normalized to $Q_{rms}=18 \mu k$.  

We generate Gaussian as well as non-Gaussian (as explained above) 
HEALPix\footnote{http://www.eso.org/kgorski/healpix/} (G\'orski, Hivon \& Wandelt 1999) projected  CMB maps that include the 
observational constraints affecting the {\it COBE}-DMR maps. We apply the
method to the 4-year
53 and 90 GHz {\it COBE}-DMR maps after adding them using inverse-noise-variance weights. These maps have a resolution of $N_{side}=32$ corresponding
to a pixel size of $\approx 110'$. Only the pixels outside the extended
Galactic cut -- described in Banday et al (1997) but explicitly recomputed for the HEALPix sky maps -- are considered in the computation. 
Best-fit monopole and dipole are removed from the pixels under analysis. 
To follow these {\it COBE}-DMR observational constraints,
the simulated Gaussian and non-Gaussian maps are convolved with a 
window function characterising the {\it COBE}-DMR one as defined in 
Wright et al. (1994).  Noise maps, generated with the same 
observation pattern and rms noise levels as the {\it COBE}-DMR data, 
are added to them. The extended Galactic cut is also impossed to
these simulated maps and a best-fit monopole and dipole is as
well removed from the pixels outside the cut.
5000 simulated maps are used in the statistical comparison described in the
next section.

\section{Method}

The aim of this work is to obtain the constraint impossed  by the {\it COBE}-DMR 
data on the nonlinear coupling parameter $f_{nl}$.
We use an approach based on the information provided by wavelets at
several scales. There are at the moment only two wavelets
implemented on the sphere that have been used in CMB analyses; the Spherical Mexican Hat wavelet 
(Cay\'on et al. 2001; Mart\'\i nez-Gonz\'alez et al. 2001) and the 
Spherical Haar wavelet (Tenorio et al. 1999; Barreiro et al. 2000). The former has proved to be 
more efficient than the Spherical Haar wavelet in detecting certain
kind of non-Gaussianity as shown in Mart\'\i nez-Gonz\'alez et al. (2001).
Moreover, the non-Gaussianity introduced in the CMB maps by the nonlinear
term added to the gravitational potential, is not expected to be 
dominated by any prefered direction. Therefore a wavelet with 
spherical symmetry, as the Spherical Mexican Hat, 
appears to be more appropriate to analyse these data.

Simulated Gaussian and non-Gaussian maps as well as the original
{\it COBE}-DMR maps are convolved with Spherical Mexican Hat wavelets 
with sizes ranging from 2 (220 arcmins) to 8 pixels. For these maps
we compute skewness and kurtosis. The non-Gaussianity introduced by
the nonlinear term added to the gravitational potential will be 
visible when we compare distributions for these statistics for the 
Gaussian and non-Gaussian cases.  Moreover, since the nonlinear term
is small in amplitude and quadratic, we will expect the third order cumulant 
to be larger than any of the higher order cumulants. 
As an example, the skewness and 
kurtosis distributions
of Gaussian and non-Gaussian maps at 220 arcmins are compared in
Figures 1 and 2. The weakly non-Gaussianity introduced in the CMB
simulations reflects in skewed distributions. As expected, the 
kurtosis seems not to be affected by the nonlinear term. 
We will therefore only use the skewness in our final analysis.

In order to combine all the information
we have at the seven scales used in the analyses we use the 
Fisher discriminant. Being $\vec x$ the 7-dimensional data vector 
including the skewness values at the seven scales, the
Fisher discriminant $t(\vec x)$ is defined as 
$$t(\vec x)=(\vec \mu_o-\vec \mu_1)^TW^{-1}\vec x.$$
Where the $o$ subscript refers to the Gaussian case and the 
$1$ subscript refers to the non-Gaussian cases. $W=V_o+V_1$ gives the 
sum of the covariance matrixes and $\vec \mu_o,\vec \mu_1$ refer
to the mean values obtained for the Gaussian and non-Gaussian case
respectively. 
This statistic has been previously used in other analysis 
related to detecting non-Gaussianity in the CMB (Barreiro \& Hobson 2001;
Mart\'\i nez-Gonz\'alez et al. 2001). 
We establish the {\it COBE}-DMR constraint on $f_{nl}$ by comparing 
the {\it COBE}-DMR Fisher discriminant value with the Fisher 
discriminant distributions
obtained for non-Gaussian simulations.

\section{Results}

\begin{figure*}
 \epsffile{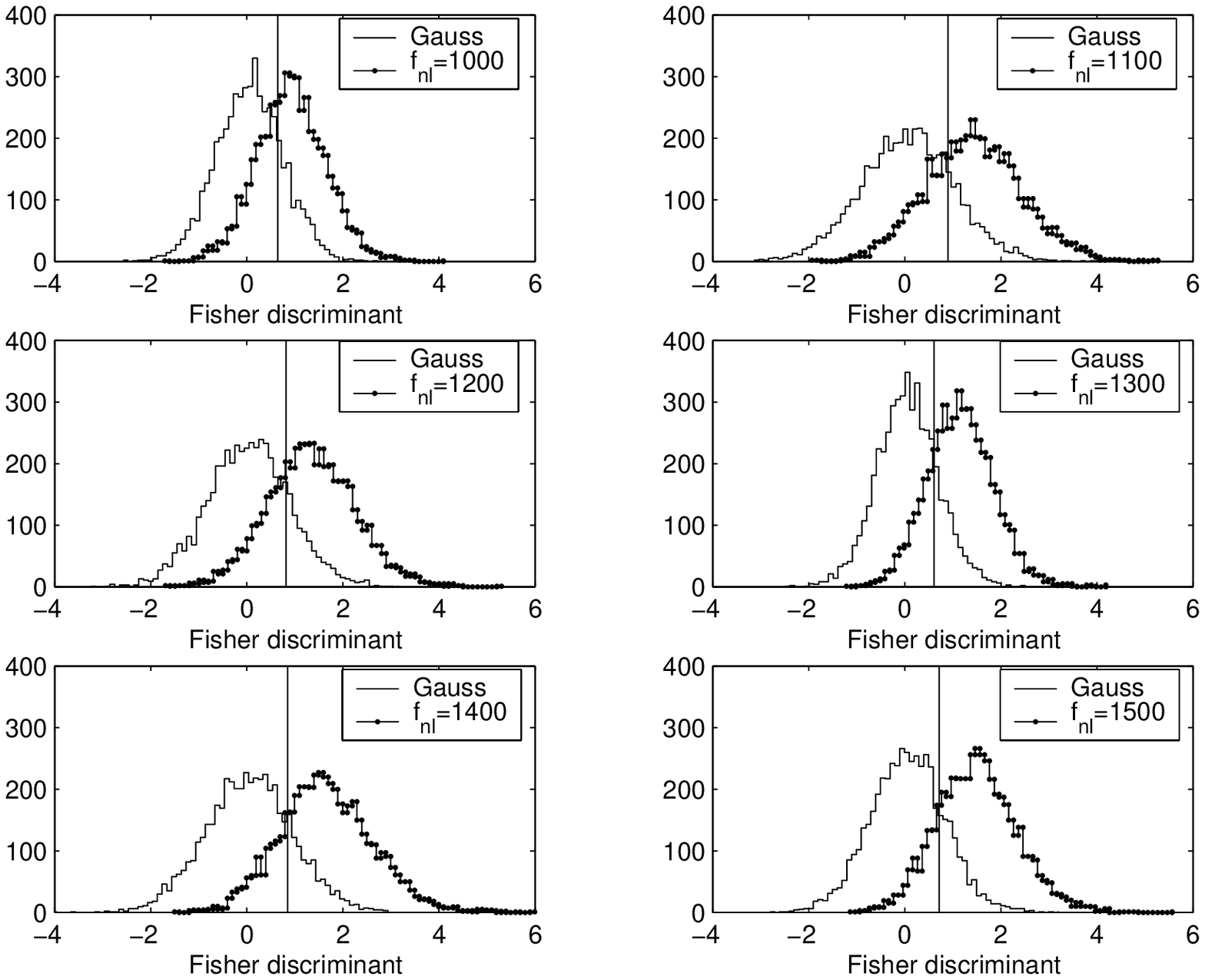}
 \caption{Distribution of Fisher discriminant values obtained from simulated Gaussian
(solid line) and NonGaussian (solid-dotted line). The non-Gaussian
simulations are generated with nonlinear coupling coefficients $f_{nl}=1000,
1100,1200,1300,1400,1500$. Vertical bars indicate the locus of the 
{\it COBE}-DMR Fisher discriminant for each case.}
 \label{f1}
\end{figure*}

Fisher discriminant distributions are presented in Figure 3 for several
non-Gaussian cases characterised by the value of the nonlinear 
coupling coefficient. Comparison of Figures 1 and 3 shows the 
expected Fisher discriminant result of increasing the power to 
distinguish two distributions.
Vertical bars in Figure 3 are drawn at the {\it COBE}-DMR Fisher discriminant
values for each non-Gaussian case considered. In order to obtain a 
constraint on $f_{nl}$ we calculate the probability of 
observing a Fisher discriminant value larger than the {\it COBE}-DMR 
one for each of the non-Gaussian cases. These probabilities are presented
in Table 1. As one can see the {\it COBE}-DMR data set a 
constraint on $f_{nl}<1100$ at the $68\%$ confidence level.

\begin{table*}
  \begin{minipage}{170mm}
  \begin{center}
  \caption[]{Probability of having non-Gaussian values of the Fisher discriminant larger(for the positive $f_{nl}$ values) or
smaller (for the negative $f_{nl}$ values) than the {\it COBE}-DMR values}
  \label{tab1}
  \begin{tabular}{c|c}\hline
  $f_{nl}$& Prob($\%$)\cr
  \hline\hline 
	
	1000& 63.0\cr
	1100& 68.7\cr
	-1100&98.9\cr
	1200& 70.2\cr
	1300& 76.6\cr
	-1300&99.4\cr
	1400& 78.6\cr
	1500& 81.2\cr
	2000& 93.4\cr 
	2500& 97.9\cr
	3000& 99.3\cr

  \hline\hline
  \end{tabular}
  \end{center}
\end{minipage}
\end{table*}

In all previous calculations we have only considered positive values
of the nonlinear coupling coefficient. To study the effect of having a 
negative $f_{nl}$ on the Fisher discriminant values, we have simulated
5000 non-Gaussian maps with $f_{nl}=-1100,-1300$. 
The distribution of 
temperature fluctuations in the map
would be expected in this case 
to be shifted 
towards negative values in the same way as the distribution of 
temperature fluctuations is shifted towards positive values
for $f_{nl}>0$ (as can be seen from the skewness distributions
presented in Figure 1). Therefore, we would expect the distribution of skewness values 
in the case of $f_{nl}<0$ to be towards the left of the 
distribution for $f_{nl}=0$. Same behaviour is observed for 
the Fisher discriminant distributions as can be seen in Figure 4.
The cases for $f_{nl}=1100,1300$ are also included in that
figure for comparison.   
The power to distinguish 
the Gaussian and non-Gaussian distributions is approximately the 
same for the positive and negative values of $f_{nl}$. At the 
$95\%$ confidence level, the power of the Fisher discriminant test
is $40.7\%, 39.1\%, 51.5\%, 48.6\%$ for $f_{nl}=1100,-1100,1300,-1300$
respectively. The same ``symmetric'' behaviour would therefore be expected for 
the any negative $f_{nl}$ value in relation to the positive one. 
As one can see from Figure 4 the locus of the {\it COBE}-DMR Fisher 
discriminant value does not change much (the probability of obtaining 
Gaussian values larger than the {\it COBE}-DMR ones is 
approximately the same).  The probability 
of having non-Gaussian values smaller than {\it COBE}-DMR ones is $98.9\%$ for
the case of $f_{nl}=-1100$ and $99.4\%$ for the case of $f_{nl}=-1300$.
The negative values of $f_{nl}$ are therefore better constrained than
the positive ones.
We can conclude that the {\it COBE}-DMR data set a 
constraint on the absolute value of the nonlinear coupling
parameter of $\vert f_{nl}\vert <1100$ at the $68\%$ confidence level.

In order to account for the possible influence of foregrounds 
we have also analysed the {\it COBE}-DMR data after foreground removal (using
{\it COBE}-DIRBE data). The results obtained are consistent with 
the ones presented above.

\begin{figure*}
 \epsffile{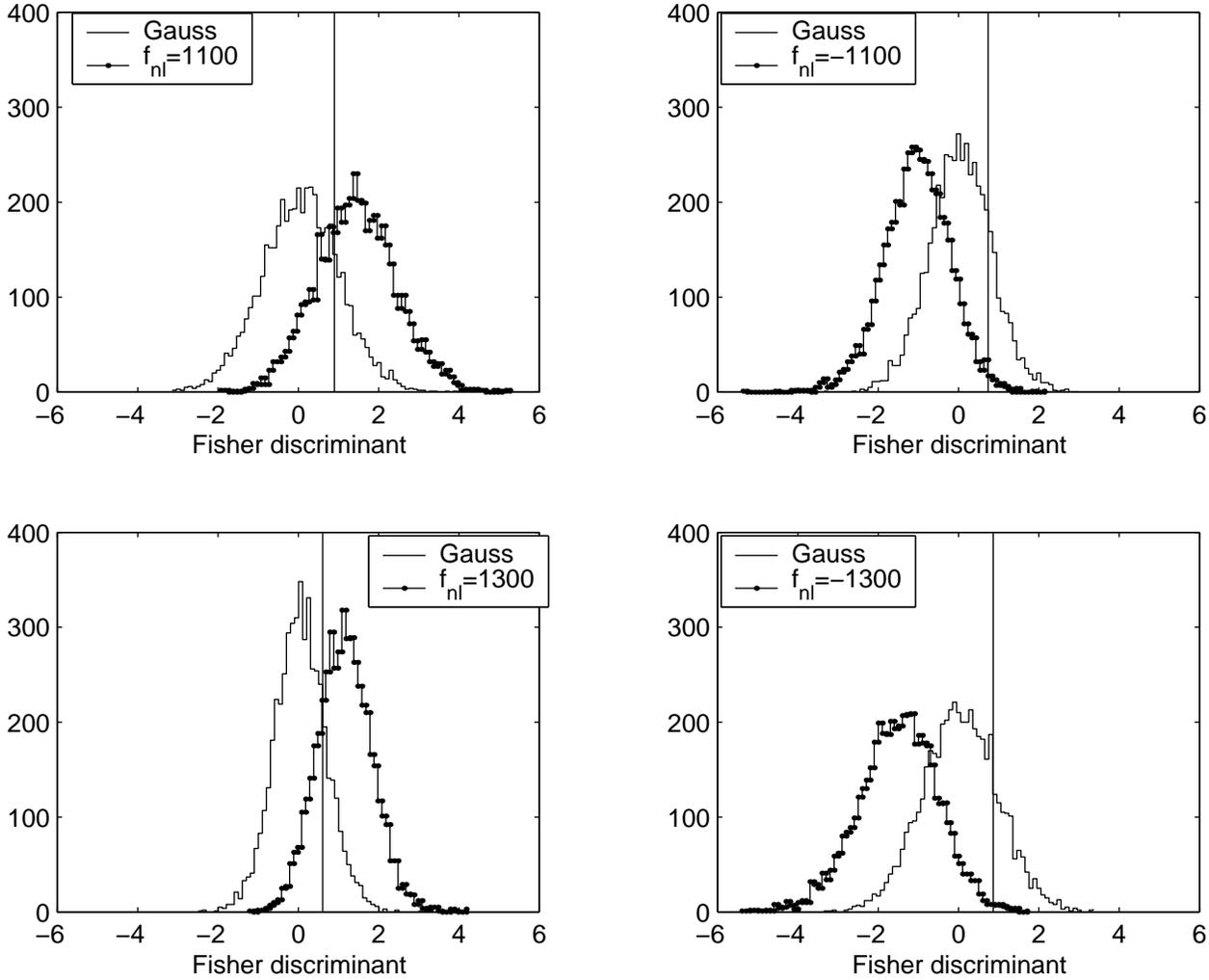}
 \caption{Distribution of Fisher discriminant values obtained from simulated Gaussian
(solid line) and NonGaussian (solid-dotted line). The non-Gaussian
simulations are generated with nonlinear coupling coefficients $f_{nl}=1100,
-1100,1300,-1300$. Vertical bars indicate the locus of the 
{\it COBE}-DMR Fisher discriminant for each case.}
 \label{f1}
\end{figure*}

\section{Discussion and Conclusions}

Weakly non-Gaussian CMB temperature fluctuations can be generated 
by the presence of a nonlinear term in the gravitational potential. 
The amount of nonlinearity can be controlled by the so called
nonlinear coupling parameter $f_{nl}$.
At large scales, the CMB temperature fluctuations are proportional
to the gravitational potential. Simulations of weakly 
non-Gaussian CMB temperature fluctuations maps at these large scales 
are therefore reasonably easy to make. 

The aim of this paper is to set a constraint on the value of the 
nonlinear coupling parameter by using the {\it COBE}-DMR data. We 
perform 5000 simulations of Gaussian and several non-Gaussian models
characterised by the value of $f_{nl}$, impossing the {\it COBE}-DMR observational
constraints. With the aid of the Spherical Mexican Hat wavelet,
we are able to obtain the skewness of these maps
at several scales. All this information is combined into the 
Fisher discriminant. Distributions of this statistic are used to set 
the constraint on $f_{nl}$ by comparing the ones obtained 
from the non-Gaussian simulations  to the {\it COBE}-DMR values. 
With this method we are able to set $\vert f_{nl}\vert <1100$ at the
$68\%$ confidence level. 
This is a tighter contraint than the one previously found by
Komatsu et al. (2002). 
They obtain the best-fit value of $f_{nl}$, by fitting the observed {\it COBE}-DMR bispectrum to a theoretical non-Gaussian model. They estimate statistical uncertainties of this parameter and place a $68\%$ confidence  limit on $f_{nl}$ as  $\vert f_{nl}\vert<1500$ for the extended cut.

A method based on wavelets in which an efficient combination of information
from different scales is used, has been shown to be very powerful in
order to test the non-Gaussianity introduced in CMB maps by a nonlinear term in the 
gravitational potential. This work has been carried out at large scales 
at which the Sachs-Wolfe effect dominates. However tighter constraints 
are expected to be obtained from the analysis of higher resolution maps.

\end{document}